\documentclass{raa}            

\usepackage{graphicx,times,natbib,url,amsmath}             

\usepackage{amsmath}

\def\lapp{\ifmmode\stackrel{<}{_{\sim}}\else$\stackrel{<}{_{\sim}}$\fi}
\def\gapp{\ifmmode\stackrel{>}{_{\sim}}\else$\stackrel{<}{_{\sim}}$\fi}
\begin{document}

   \title{The Role of FAST in Pulsar Timing Arrays
}

   \volnopage{Vol.0 (200x) No.0, 000--000}      
   \setcounter{page}{1}          

   \author{George Hobbs
      \inst{1}
   \and Shi Dai
       \inst{2,1}
   \and Richard N. Manchester
      \inst{1}
         \and Ryan M. Shannon
      \inst{1}
             \and Matthew Kerr            
      \inst{1}
            \and Kejia Lee
       \inst{3,4}
         \and Renxin Xu
      \inst{2,3}
   }

   \institute{ CSIRO Astronomy and Space Science, PO Box 76, Epping, NSW 1710, Australia; {\it george.hobbs@csiro.au}\\
        \and
             Department of Astronomy, School of Physics, Peking University, Beijing 100871, China \\
       \and
           Kavli Institute for Astronomy and Astrophysics, Peking University, Beijing 100871, China \\
       \and
           Max-Planck-Institut f{\"u}r Radioastronomie, Auf Dem H{\"u}gel 69, Bonn, 53121, Germany \\
   }

   \date{Received~~2013 month day; accepted~~2013~~month day}

\abstract{ 
The Five-hundred-meter Aperture Spherical Telescope (FAST) will become one of the world-leading telescopes for pulsar timing array (PTA) research. The primary goals for PTAs are to detect (and subsequently study) ultra-low-frequency gravitational waves, to develop a pulsar-based time standard and to improve solar system planetary ephemerides.  FAST will have the sensitivity to observe known pulsars with significantly improved signal-to-noise ratios and will discover a large number of currently unknown pulsars.  We describe how FAST will contribute to PTA research and show that jitter- and timing-noise will be the limiting noise processes for FAST data sets.  Jitter noise will limit the timing precision achievable over data spans of a few years while timing noise will limit the precision achievable over many years.
}

   \authorrunning{G. Hobbs, et al.}            
   \titlerunning{The Role of FAST in Pulsar Timing Arrays}  

   \maketitle

%
%
\section{Introduction}           
\label{sect:intro}

With its massive collecting area, the Five-hundred-meter Aperture Spherical Telescope (FAST) is expected to revolutionise pulsar astronomy.  In terms of physical size, the closest existing single dish is the 300\,m Arecibo radio telescope. This telescope is renowned for detecting the first planets outside the solar system (around PSR B1257$+$12; Wolszczan \& Frail 1992\nocite{wf92}) and the discovery and analysis of the first pulsar in a binary system \citep{ht75} which provided stringent tests on the general theory of relativity (and led to the Nobel Prize in physics).   These results were based on measuring pulse times-of-arrival (ToAs) over many years. In the pulsar timing technique the ToAs are compared with model predictions using a model that describes the position, orbit and spin-down of the pulsar (see e.g., Edwards, Hobbs \& Manchester 2006\nocite{ehm06}).   The differences between the predictions and the measured arrival times are the timing residuals and can be used to identify phenomena that affect the pulse arrival times but are either not included in the model, or not included with sufficient precision. The timing model is iteratively updated in order to minimise the residuals.  In this paper we describe the search for phenomena that affect the pulse ToAs at the 10--100\,ns level, such as the effects of ultra-low frequency gravitational waves passing through the solar neighbourhood.  For most pulsars, the precision with which the ToAs can be determined is not sufficient to search for such small effects.  However, observations with the Arecibo telescope also led to the discovery of millisecond pulsars \citep{bkh+82}.  The rotation of these pulsars is extremely stable and current observations of some of these millisecond pulsars are leading to data sets with root-mean-square (rms) timing residuals of $<100$\,ns.

Pulsar timing experiments do not require a large field-of-view, but because pulsars are generally faint radio sources ($\sim$\,mJy) they do require a very sensitive telescope.  The use of an interferometer adds a significant level of complexity and so the ideal pulsar telescope consists of a single antenna with a large collecting area. FAST fits these criteria and therefore should become a world-leading telescope for pulsar timing experiments.  

\begin{figure}
\includegraphics[width=5.5cm,angle=-90]{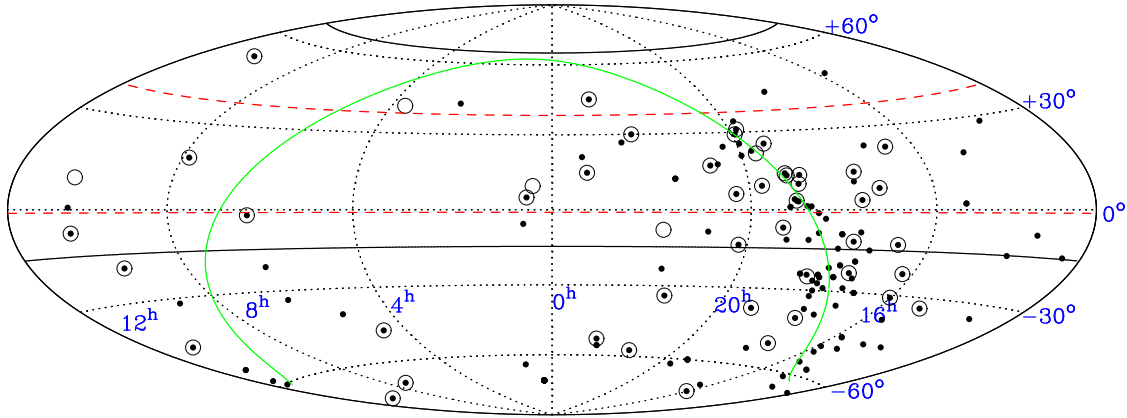}
\caption{Sky map in equatorial coordinates containing the IPTA pulsars (open circles) and all known millisecond pulsars that have pulse periods $P < 15$\,ms and $\dot{P} < 10^{-19}$ (dots).  The black solid lines indicate the declination limits for FAST. The red dashed lines give the declination limits for Arecibo. The green solid line indicates the Galactic plane.}\label{fg:sky}
\end{figure}

Pulsar Timing Array (PTA) projects differ from more traditional pulsar timing experiments in that they aim to extract common signals present within the timing residuals for multiple pulsars. Residuals caused by the irregular rotation of a given pulsar, the interstellar medium or insufficient precision in any of the pulsar parameters will be uncorrelated between pulsars. In contrast, an error in the terrestrial time standard used for measuring the pulse arrival times would lead to an identical signal in the timing residuals of all pulsars.  The pulsar timing method relies on converting the pulse arrival times to the solar system barycentre (SSB; Edwards et al. 2006). Errors in the position of the Earth with respect to the SSB will lead to timing residuals that are not identical between different pulsars, but are correlated (the effect will depend on a pulsar's ecliptic latitude). Gravitational waves (GWs) passing the Earth will lead to variations that exhibit a signal that depends upon the pulsar-Earth-GW angle. 

Even though the basic concepts of a PTA were laid out by \cite{fb90} it was not until 2004 when the first major PTA (the Parkes Pulsar Timing Array; PPTA) started observing enough pulsars with sufficient sensitivity to have a chance to succeed in the PTA goals.  The PPTA currently observes 24 pulsars using the 64-m diameter Parkes radio telescope in Australia (see Manchester et al. 2013\nocite{mhb+13} and Hobbs 2013\nocite{hob13} for recent reviews).   The North American PTA (the North American NanoHertz Observatory for GWs; NANOGrav), described by \cite{dfg+13} and \cite{mcl13}, formed in October 2007 and carries out observations with the Arecibo and Green Bank telescopes.   The European PTA (EPTA; Kramer \& Champion 2013\nocite{kc13}) was established in 2004/2005 and includes telescopes in England, France, Germany, the Netherlands and Italy.   During 2008 an agreement was made to share data sets between the three major PTAs. This led to the formation of the International Pulsar Timing Array (IPTA; see Hobbs et al. 2010\nocite{haa+10}, Manchester 2013\nocite{man13} and references therein).

In this paper, we first describe the FAST telescope (\S~\ref{sec:FAST}). We then summarise the status of current PTA projects (\S~\ref{sec:currentStatus}).   In \S~\ref{sec:whatNext} we describe how FAST is likely to contribute to the IPTA.  In \S~\ref{sec:limitNoise} we highlight noise processes likely to dominate the resulting data sets.  \S~\ref{sec:realistic} contains a description of a realistic array that could be observed using FAST and describes requirements relating to data archiving.  Finally, we describe some key unanswered questions relating to PTAs and FAST (\S~\ref{sec:unanswered}).

\section{The FAST telescope}\label{sec:FAST}

The FAST telescope has been described in \cite{nlj+11} with updates in \cite{lnp13}. It is expected to conduct initial observations by September 2016. The telescope is situated in Guizhou province in China at a longitude of 107$^\circ$21$^\prime$ and latitude of $+25^\circ48^\prime$ .  Initially it will have a maximum zenith angle of 40$^\circ$ providing an observable declination range of $-15^\circ < \delta < 65^\circ$.  The telescope has an effective diameter of 300\,m and a maximum slew time between sources of 10\,min.  It is likely that initial PTA observations on FAST will use a single pixel wide-band receiver spanning from $\sim 300$\,MHz to $1.7$\,GHz \citep{lnp13}.  The full suite of receivers is not expected for at least one extra year.  After that time PTA research will also make use of a receiver covering a band from 2 to 3\,GHz. 

In Figure~\ref{fg:sky} we provide a map of the sky in equatorial coordinates.  The FAST declination range is indicated by solid black lines. All the currently known millisecond pulsars that may be suitable for PTA work are shown as small dots.  Those currently being observed as part of the IPTA project \citep{man13} are indicated by open circles.  As listed in Table~\ref{tb:fastPsrs}, FAST will therefore be able to observe 32 out of the 50 pulsars currently being observed for PTA purposes.  The first five columns of the table provide the pulsar's name, pulse period, dispersion measure, pulse width and the flux density in the 20\,cm band. These parameters were obtained from the ATNF pulsar catalogue \citep{mhth05}. The remaining columns in this table are described later.

\begin{table}
\caption{Pulsars Detectable by FAST and Currently Observed as Part of the International Pulsar Timing Array}\label{tb:fastPsrs}
\begin{tabular}{llrllllllll}
\hline
PSR J & P & DM & W$_{\rm 50}$ & S$_{1400}$ & Jitter & Jitter & Jitter & T$_{100 {\rm ns}}$ & T$_{30 {\rm ns}}$ & $\sigma_{\rm 15 min}$\\
            &     &        &                            &                       & Dom.? & Dom.? & Dom.? & (min) & (hr)  & ($\mu$s)  \\
            & (ms) & (cm$^{-3}$pc) & (ms) & (mJy) & (FAST) & (Parkes)  & (Qitai) & (FAST) & (FAST) & (FAST)\\
\hline
J0023$+$0923 & 3.1 & 14.3 & -- & -- \\
J0030$+$0451 & 4.9 & 4.3 & -- & 0.6  \\
J0340$+$4130 & 3.3 & 49.6 & -- & -- \\
J0613$-$0200 & 3.1 & 38.8 & 0.5 & 2.3 & Y & N & N & 52 & 9.6 & 0.19 \\
J0751$+$1807 & 3.5 & 30.3 & 0.7 & 3.2 & Y & N & N & 114 & 21.1 & 0.28 \\
\\
J1012$+$5307 & 5.3 & 9.0 & 0.7 & 3.0 & Y & N  & N & 173 & 32.1 & 0.34 \\
J1022$+$1001 & 16.5 & 10.3 & 1.0 &  6.1 & Y & N & Y & 1100 & 200 & 0.86 \\
J1024$-$0719 & 5.2 & 6.5 & 0.5 & 1.5 & Y & N & N & 87 & 16 & 0.24 \\
J1640$+$2224 & 3.2 & 18.4 & 0.2 & 2.0 & Y & N & N & 9 & 1.6 & 0.086 \\
J1643$-$1224 & 4.6 & 62.4 & 0.3 & 4.8 & Y & N & Y & 28 & 5.1 & 0.14 \\
\\
J1713$+$0747 & 4.6 & 16.0 & 0.1 & 10.2 & Y & N & Y & 3.1 & 0.6 & 0.045 \\
J1738$+$0333 & 5.9 & 33.8 & 0.4 & -- & \\
J1741$+$1351 & 3.7 & 24.2 & 0.2 & 0.9 & Y & N & N & 9.9 & 1.8 & 0.081  \\
J1744$-$1134 & 4.1 & 3.14 & 0.1 & 3.1 & Y & N & Y & 2.7 & 0.5 & 0.042 \\
J1853$+$1303 & 4.1 & 30.6 & -- & 0.4 \\
\\
J1857$+$0943 & 5.4 & 13.3 & 0.5 & 5.0 & Y & N & Y & 90 & 17 & 0.24 \\
J1903$+$0327 & 2.1 & 297.5 & -- & 1.3 \\
J1910$+$1256 & 5.0 & 38.1 & -- & 0.5 \\
J1911$+$1347 & 4.6 & 31.0 & 0.2 & 0.1 & N & N & N & 278 & 52 & 0.43 \\
J1918$-$0642 & 7.6 & 26.6 & 0.7 & 0.6 & Y & N & N & 248 & 46 & 0.41 \\
\\
J1923$+$2515 & 3.8 & 18.9 & 0.5 &  -- \\
J1939$+$2134 & 1.6 & 71.0 & 0.04 & 13.2 & Y & N & Y & 0.2 & 0.03 & 0.010 \\
J1944$+$0907 &  5.2 & 24.3 & 0.5 & -- \\
J1949$+$3106 & 13.1 & 164.1 & -- & 0.2 \\
J1955$+$2908 & 6.1 & 104.5 & 1.8 & 1.1 & N  & N & N & 1715 & 318 & 1.06 \\
\\
J2010$-$1323 & 5.2 & 22.2 & 0.3 & 1.6 & Y & N & N & 31.2 & 5.8 & 0.14 \\
J2017$+$0603 & 2.9 & 23.9 & -- & 0.5 \\
J2043$+$1711 & 2.4 & 20.7 & -- & -- \\
J2145$-$0750 & 16.1 & 9.0 & 0.3 & 8.9 & Y  & Y & Y & 97 & 17.9 & 0.25 \\
J2214$+$3000 & 3.1 & 22.6 & -- & -- \\
\\
J2302$+$4442 & 5.2 & 13.7 & -- & 1.2 \\
J2317$+$1439 & 3.4 & 21.9 & 0.5 & 4.0 & Y & N & N & 57 & 10.4 & 0.19 \\
\hline

\end{tabular}
\end{table}

\section{How FAST will contribute to PTA goals}\label{sec:currentStatus}

We begin this section by describing the current status of the science goals achievable with a PTA and then describe how FAST will contribute to those goals.   The primary goals are to:

\begin{itemize}
\item {\emph{Develop a pulsar-based time standard:} \cite{hcm+12} showed how signals common to all pulsars could be identified in a given PTA data set. This method was applied to PPTA data sets and the known offsets between the world's best realisations of Terrestrial Time (TT) were recovered (for this work TT as realised by International Atomic Time, TT(TAI), and the post-corrected realisations from the Bureau International des Poids et  Mesures, TT(BIPM), were compared).  This work therefore demonstrated that a pulsar based timescale can be constructed. The published timescale will be significantly improved by the addition of observations from the EPTA and NANOGrav.  The development of a time scale based on IPTA data is currently underway.}
\item {\emph{Improve the solar system ephemeris:} An error in the assumed vector between the observatory and the solar system barycentre (SSB) leads to induced timing residuals that depends upon the direction and size of the error in the vector and the direction to the pulsar.  Different pulsars will therefore exhibit different residuals and, with a sufficiently large number of pulsars, the cause of such residuals can be identified. \cite{chm+10} searched specifically for the signatures of incorrect mass estimates of the planetary systems in our solar system and succeeded in publishing a precise determination of the mass of the Jovian system.  This work is now being updated and continued as part of an IPTA project. Software is currently being extended to enable a search for an unknown mass in the solar system by directly fitting for the components of any error in the observatory--SSB vector.}
\item {\emph{Detect nanohertz-frequency GWs:} There are numerous GW sources that could lead to detectable signatures in pulsar data sets.  These include individual sources, bursts with memory and stochastic backgrounds. 

A single (effectively non-evolving) binary system of supermassive binary black holes produces a sinusoidal signal.  Various algorithms have recently been developed to search for such waves.  These algorithms include frequentist-based approaches (Jenet et al. 2004\nocite{jllw04}, Yardley et al. 2010\nocite{yhj+10} and Zhu et al., submitted) and Bayesian methods (Lee et al., 2011\nocite{lwk+11}, Babak \& Sesana 2012\nocite{bs12}, Ellis, Siemens \& Creighton 2012\nocite{esc12}, Petiteau et al. 2013\nocite{pbs+13}, Ellis 2013\nocite{ell13}).  

The merger of two supermassive black holes may lead to a detectable memory event that takes the form of a ``glitch'' in the timing residuals (van Haasteren \& Levin 2010\nocite{vl10}, Cordes \& Jenet 2012\nocite{cj12}, Wang et al., submitted).  

Most recent work has concentrated upon searching for a GW background formed by the superposition of a large number of GWs from black hole binaries (see e.g., Sesana, Vecchio \& Colacino 2008\nocite{svc08}, Ravi et al. 2012\nocite{rwh+12}) as this is predicted to be the dominant signal in the PTA band. Background from cosmic (super)strings (e.g., {\"O}lmez, Mandic \& Siemens 2010\nocite{oms10}) and inflation (e.g., Tong et al. 2014\nocite{tzz14}) have also been studied.  Various algorithms have been published to constrain or to detect GW backgrounds (see e.g., Jenet et al. 2006\nocite{jhv+06}, Yardley et al. 2011\nocite{ych+11}, van Haasteren et al. 2011\nocite{vlj+11}, Sanidas, Battye \& Stappers 2012\nocite{sbs12}, Demorest et al. 2013\nocite{dfg+13}).  \cite{src+13} recently used PPTA data sets to obtain the most stringent constraint to date on the GW background.}
\end{itemize}

The high-precision long data sets on a large number of some of the most extreme pulsars clearly leads to numerous other opportunities.  These include (but are not limited to) studying
\begin{itemize}
\item \emph{the interstellar medium:} Studies of the time variability of pulsar dispersion measures (DMs; the integrated electron density along the line of sight to the pulsar) probe variations in the interstellar medium.  PTA-quality data sets allow precise DMs to be determined with a high observing cadence.  Using PPTA observations, \cite{yhc+07a} demonstrated that the DM variations for some pulsars did not follow the predictions expected from Kolmogorov turbulence.  This work was continued by \cite{kcs+13} who identified, for some pulsars, a wavelength-dependent annual variation that was explained as a persistent gradient of electron density on an astronomical-unit spatial scale.  

 Observations of various PTA pulsars (PSRs J1939$+$2139, J1643$-$1224 and J1603$-$7202 reported in Cognard et al. 1993\nocite{cbl+93}, Maitia, Lestrade \& Cognard 2003 and Keith et al. 2013\nocite{kcs+13} respectively) have been used to detect extreme scattering events.  \cite{sti13} and Demorest (2011) have described research relating to correcting for multi-path scattering effects using phase reconstruction techniques and cyclic spectroscopy.

\item \emph{the solar wind: } \cite{ychm12} showed that it was possible to measure the path-integrated electron density and the Faraday rotation simultaneously at small radii from the Sun using PTA pulsars as a linearly polarised radio source.

\item \emph{properties of individual pulsars:} Even though the PTA goals are related to identifying correlated signals between different pulsars, many of the pulsars that make up a PTA are intrinsically interesting both as individual objects and also to provide information on the millisecond pulsar population.  For example, \cite{ymc+11} presented polarisation profiles for 20 millisecond pulsars and showed that 13 pulsars in the sample showed emission over more than half of the pulse period.  This work was continued in \cite{ymh+11} to study variations in the pulsar rotation measures.  Very little long-term variation in the interstellar rotation measure was found.

The timing solutions for most PTA pulsars contain determinations of parallax and/or the rate of change of the orbital period derivative.  Both of these can be used to obtain the distance to the pulsar.  Measuring a pulsar's distance to within a gravitational wave wavelength has not yet been achieved, but would significantly improve the sensitivity of that pulsar to gravitational wave signals (see e.g., Li et al. 2011\nocite{lwk+11}). 

\item \emph{pulsar-based navigation techniques:} One interesting side-project for PTA data sets is the possibility of using observations of millisecond pulsars to navigate spacecraft travelling through the solar system (or even beyond).  The basic algorithms have been presented by numerous authors.  \cite{dhy+13} recently used actual PPTA observations to demonstrate the effectiveness of such methods. 

\item \emph{tests of theories of gravity:} Many of the PTA pulsars are in binary systems.  The majority of such pulsars have a white dwarf companion and are not as relativistic as the neutron star-neutron star binaries.  They have, however, been used to test theories of gravity. Examples include PSR J1738$+$0333, which currently provides the most stringent test of scalar-tensor gravity \citep{fwe+12} and PSR J0437$-$4715 which was used to place a constraint on the variation of Newton's gravitational constant \citep{vbv+08}.

\end{itemize}

\subsection{What still needs to be done}\label{sec:whatNext}

We have not yet detected GWs, nor have we identified any problems with the world's best time scale TT(BIPM).  We have published a precise mass estimate for the Jovian system, but measurements from the \emph{Galileo} spacecraft are consistent with our result and are more precise (by a factor of $\sim$20).  All the main goals of the existing PTA experiments are therefore still extant.  Here we discuss the data sets that are required for us to make significant progress towards these goals.  

\subsubsection{Pulsar-based time standards}

\begin{figure}
\includegraphics[angle=-90,width=7cm]{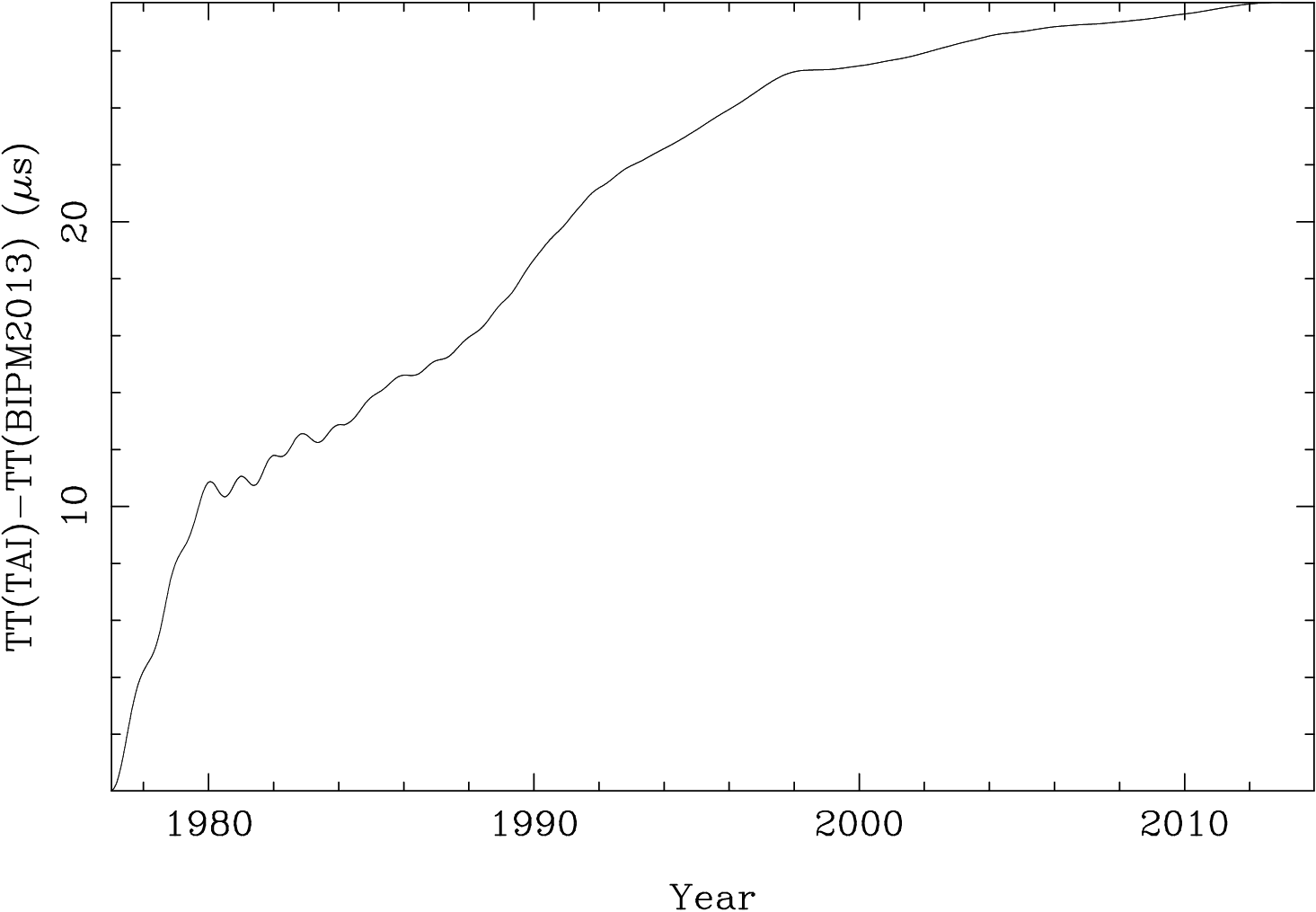}
\includegraphics[angle=-90,width=7cm]{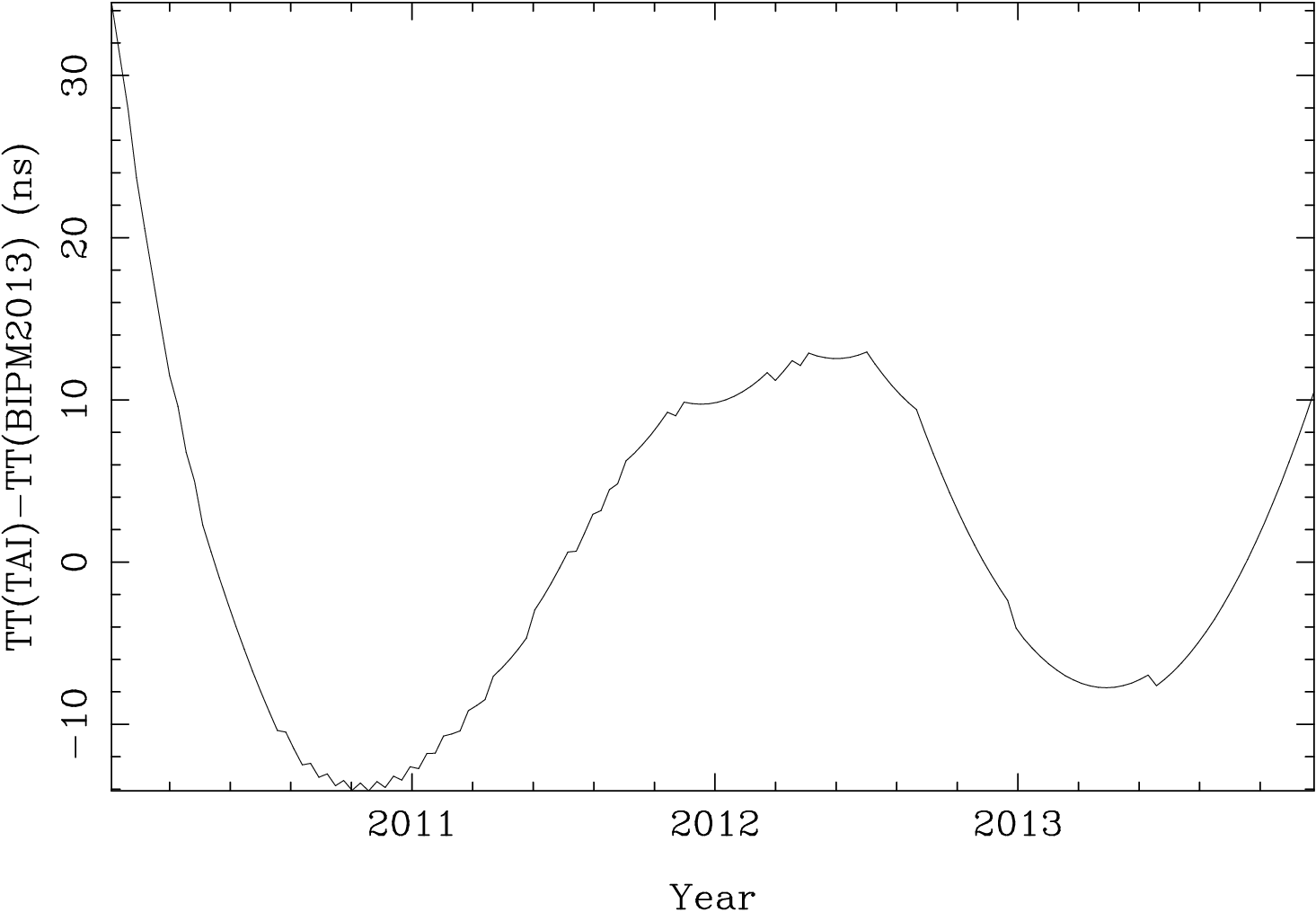}
\caption{(left) The difference between TT(TAI) and TT(BIMP2013) over the last $\sim$30\,years.  (Right) the same, but from the year 2010 and after a quadratic polynomial has been fitted and removed (see main text).} \label{fg:timeDiff}
\end{figure}

In Figure~\ref{fg:timeDiff} we show the difference between TT(BIPM2013) and TT(TAI) over the past 30 years (left panel) and since the year 2010 (right panel).  The largest component of the signal seen in the left-hand panel is a linear trend. As described by e.g., \cite{hcm+12}, it will not be possible to detect any irregularities in a terrestrial timescale that take the form of a quadratic polynomial.  In the right-hand panel of Figure~\ref{fg:timeDiff} we have therefore fitted and removed a quadratic polynomial  that has been weighted appropriately to account for the different data sets used.  In order to detect the recent irregularities in TT(TAI) the common signal in the timing residuals of all pulsars will need to be determined with a precision of $\sim 10$\,ns.  TT(BIPM2013) should be significantly more stable than TT(TAI) and so it will be necessary to measure the common signal at the nanosecond level over many years in order to identify irregularities in the world's best terrestrial timescale.  For idealised pulsar data sets that have the same sampling, timing model parameters and data spans, the pulsar time scale is obtained simply from a weighted average of the timing residuals within a given observing period.  A timing array consisting of 50 pulsars each with a timing precision of 100\,ns would therefore allow the pulsar timescale to be determined at the 14\,ns level. For 50 pulsars each with a timing precision of 50\,ns this would reduce to $\sim$7\,ns.  

Of course,  the terrestrial time standard community will continue to improve the stability of their time scales and therefore the main use of a pulsar-based time scale will be to provide an independent check over long time spans. As emphasised by Hobbs et al. (2012) a pulsar-based time scale provides 1) a time scale based on macroscopic objects of stellar mass instead of being based on atomic clocks and 2) a time scale that is continuous and will remain valid far longer than any clock that we can construct.

\subsubsection{The solar system}

\begin{figure}
\includegraphics[angle=-90,width=7cm]{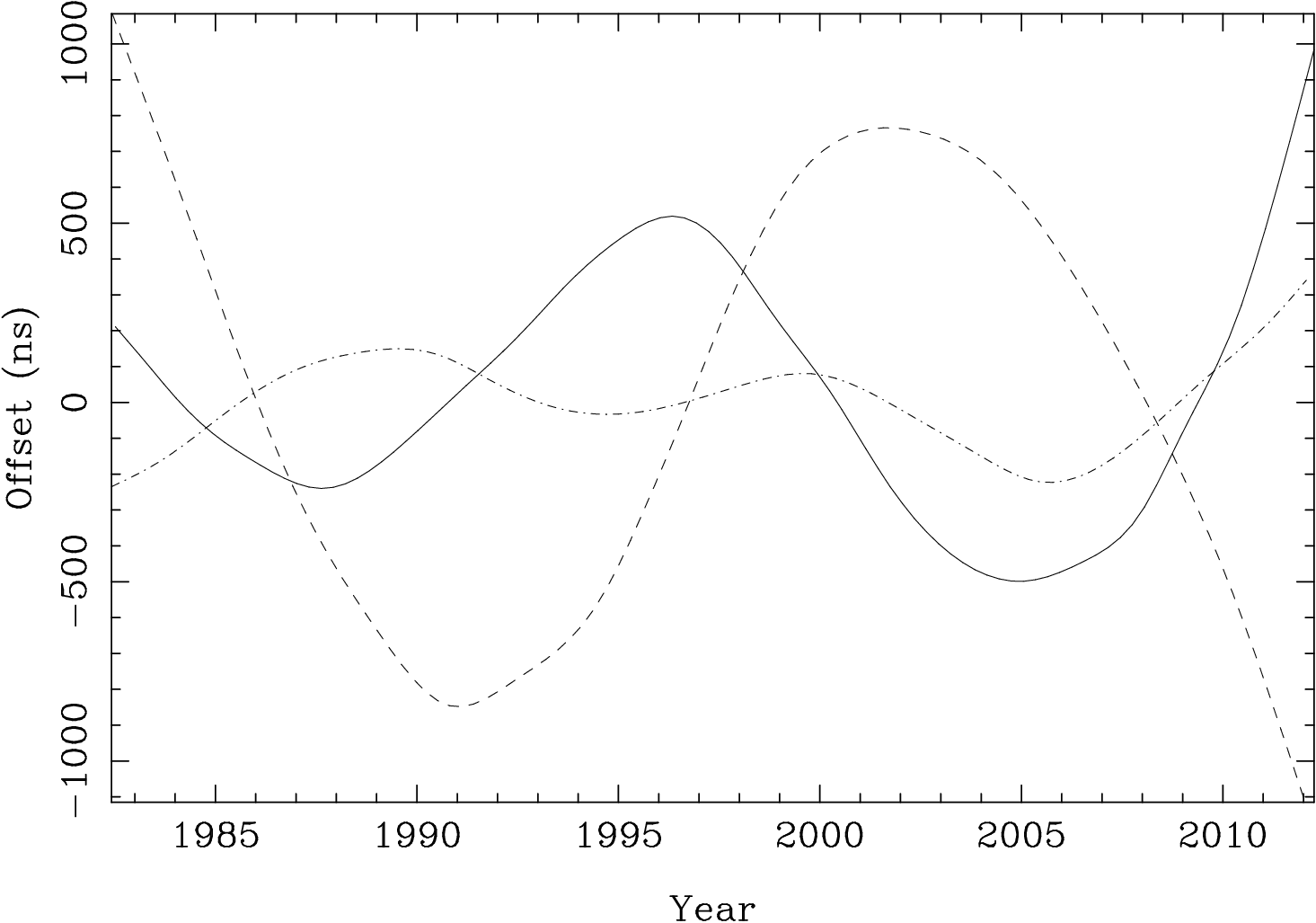}
\includegraphics[angle=-90,width=7cm]{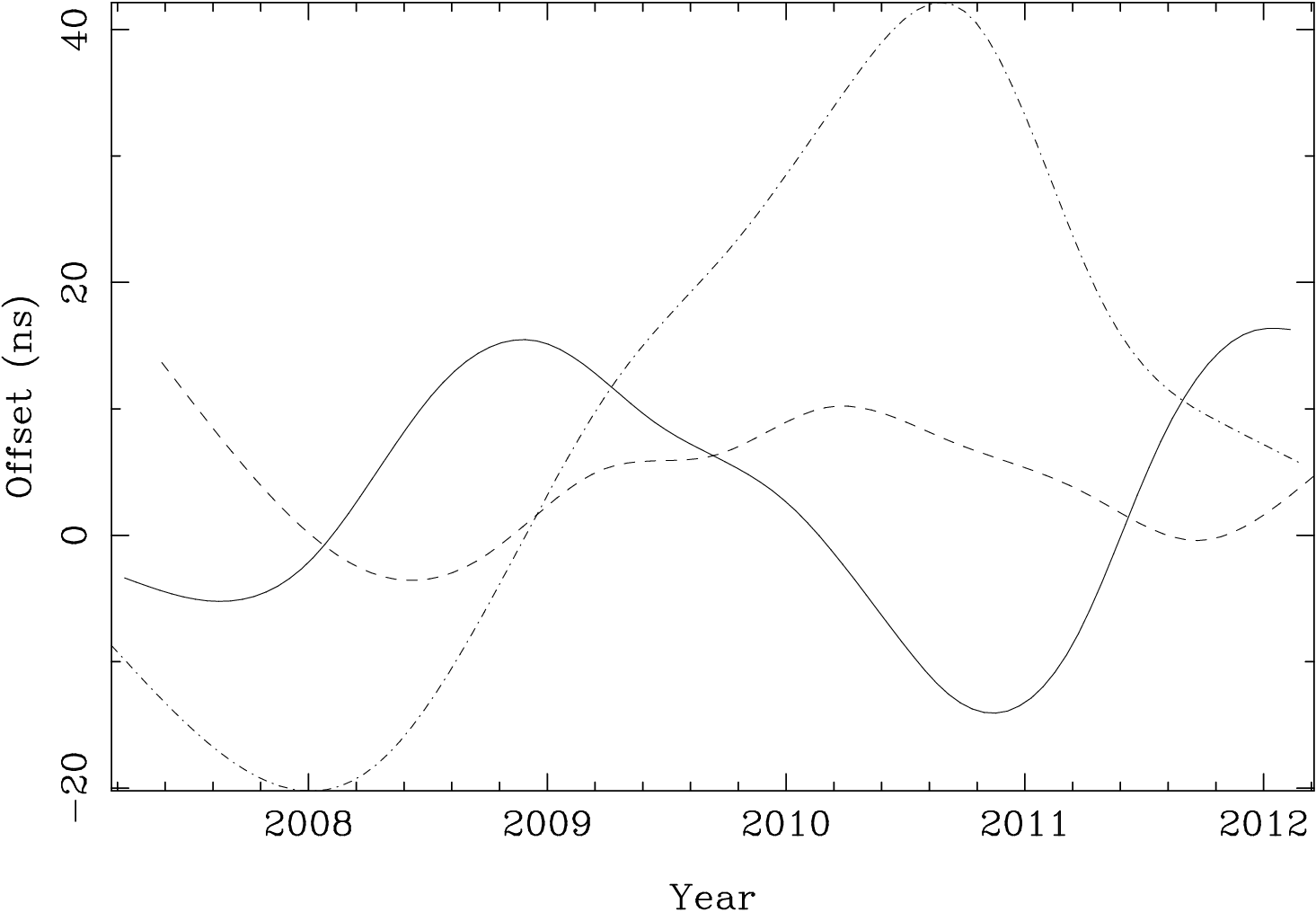}
\caption{Differences in the Earth-SSB vector between DE421 and DE414 over the last 30 years (left panel) and over 5 years (right panel). A quadratic polynomial and annual terms have been fitted and removed in both panels. The three components of the Earth-SSB vector are shown as a solid line ($\Delta$X), dashed line ($\Delta$Y) and dot-dashed line ($\Delta$Z) respectively.} \label{fg:ephDiff}
\end{figure}

PTAs are sensitive to any errors in the assumed vector from the Earth to the SSB. Three main solar system ephemerides now exist from North America (the DE series; Newhall, Standish \& Williams 1983\nocite{nsw83}), Europe (the INPOP series; Fienga et al. 2008\nocite{fmlg08}) and Russia (the EPM series; Pitjeva 2005\nocite{pit05}).  A comparison of DE421, EPM2008 and INPOP08 has been published by Hilton \& Hohenkerk (2010)\footnote{\url{http://syrte.obspm.fr/jsr/journees2010/pdf/Hilton.pdf}}.  They showed that the largest difference in the models was the treatment of small bodies such as asteroids and trans-Neptunian objects.   The EPM2008 barycentric position was shown to be significantly offset from the other ephemerides. This is caused by the inclusion of extra trans-Neptunian objects in EPM2008.  The motions of such objects are slow and, for PTA data spans, this represents a fixed offset in the position of the barycentre with respect to the other ephemerides.

The pulsar timing method cannot detect any error in the Earth--SSB vector that is modelled as a simple spatial offset or follows a quadratic variation with time.  It is therefore likely that PTAs will continue to improve mass estimates of planetary systems with orbital periods that are shorter than the PTA data spans, but will not provide strong constraints (at least via this method) on trans-Neptunian objects.  Note that this could be improved if the pulsar positions were determined accurately using a non-timing method.  For instance, Very Long Baseline Interferometry (VLBI) can be used to obtain pulsar positions.  However, making use of VLBI positions for this work requires that the reference-frame tie between the planetary ephemeris and the International Celestial Reference Frame be known (see, Madison, Chatterjee \& Cordes 2013).

In order to obtain an estimate of the errors that we may be expected, we compare the three components of the Earth--SSB vector over the past 30 years as measured using the DE421 and DE414 ephemerides.   The difference between the two ephemerides is shown in Figure~\ref{fg:ephDiff}.  Over a 30 year timescale the variations (even after removing a quadratic polynomial) are relatively large and extend over a few microseconds (Note that the induced residuals caused by such an error in the ephemeris can be calculated by forming the dot product of the vector pointing towards the pulsar and these difference components). Variations are seen on the orbital period of Jupiter (12 years) and of Saturn (29 years).  Over a much shorter time interval (right panel) the variations are at a much smaller level ($\sim 20$\,ns) and will require a large number of very precisely timed pulsars for detection.

\subsubsection{Gravitational waves}

In contrast to the aforementioned goals, the detection of ultra-low frequency GWs could, in principal, be obtained with relatively short ($\sim 5$\,yr) data sets if a large number of stable pulsars were timed with high precision.  Theoretical predictions currently suggest that the most likely GW signal will be from a stochastic background of supermassive binary black holes.  The power spectral density of the residuals induced by such a signal can be expressed as:
\begin{equation}\label{eqn:psd}
P(f) = \frac{A^2}{12\pi^2}\left(\frac{f}{f_{1yr}}\right)^{-13/3}
\end{equation}
where $f_{1yr} = 1/{1 {\rm yr}}$. The GW background amplitude, $A$, has been limited by \cite{src+13} to be $A < 2.7 \times 10^{-15}$ at 95\% confidence and is expected to lie within $10^{-16} < A < 10^{-15}$ (see Sesana 2013 and references therein\nocite{ses13}).  

A single pulsar can be used to provide an upper bound on $A$. However, to \emph{detect} a GW background the expected correlation \citep{hd83} between the timing residuals for different pulsars will need to be identified.  Determining exactly when a particular PTA will be able to make a GW detection is extremely complicated and depends upon the number of pulsars (which changes as pulsars get added or removed from an array), data spans (some pulsars have long data spans, others do not), sampling cadence and the noise processes affecting the residuals. 

Research is ongoing to make reasonable predictions for GW detection for a specified real (or future) data set.    An initial attempt has been recently published by \cite{sejr13} who calculated the time to detection for idealised PTAs.  In their work they consider three regimes: 1) where the GW signal is weak and the white noise dominates, 2) where the GW signal is strong and 3) an intermediate regime where only the power in the lowest frequencies of the GW background is above the white noise level.  They demonstrate that the significance of a detection of a GW signal will increase with longer data spans at a different rate for these three scenarios.  A realistic array is likely to be complex and contain some pulsars for which the GW signal is weak and other pulsars for which the signal will be strong.  This implies that it is very challenging to make analytic predictions for the time to detection for a given array.

Various updates to the \textsc{tempo2} software package \citep{hem06} are currently being made to allow realistic data sets to be simulated.  Such data sets can include earlier known data and predictions for future observations. The effects of various noise processes, including a GW background can then be added. GW background detection algorithms can subsequently be run on these mock data sets and realistic expectations for the time to GW detection can be produced.  This work is not yet complete and we leave to a later paper an attempt to predict when FAST, along with earlier IPTA data and combined with SKA data would be expected to detect a GW background. 

Before FAST begins observing it is possible that an initial detection of ultra-low frequency GWs would already have been made using IPTA data sets. If true, FAST would then begin the exciting work of studying those GWs.  This would include
\begin{itemize}
\item \emph{Confirming the detection:}  The most likely detection is of a GW background.  As data sets improve and get longer then the significance of the detection will increase.  FAST data sets would be essential to confirm that the signal detected is not a strange artefact related to a few specific observatories.
\item \emph{Confirming the source of the background:} The induced residuals from a background of cosmic strings, the inflationary era and merging supermassive black holes are all similar. The main difference is a different power law exponent in the characteristic strain spectrum.  From an initial detection it is likely to be challenging to distinguish between these models.  \cite{ses13} emphasises that a confirmed GW background from black hole mergers would provide ``direct unquestionable evidence of the existance of a large population of sub-parsec supermassive black hole binaries, proving another crucial prediction of the hierarchical model of structure formation".
\item \emph{Looking for a turn-over in the GW spectrum:} for a background of black holes, Equation~\ref{eqn:psd} is only valid for binary black holes in circular orbits and driven solely by GW emission (see Ravi et al., 2014\nocite{rws+14} and Sesana 2013\nocite{ses13}).   Any turn-over would indicate interactions between the binary black holes and their environment (for instance, the effects of stellar scattering or if the black holes are surrounded by circumbinary discs). \cite{sbs13} considered the GW spectrum for a GW background formed from cosmic strings and showed that it  also can have a complicated shape in the region of interest for PTA experiments.
\item \emph{Searching for anisotropies in the background:} \cite{msmv13} argue that an anisotropy may be present in a GW background formed from black hole binary systems.  They demonstrate the means by which such an anisotropy could be identified with a sufficiently sensitive PTA.
\item \emph{Testing the predictions of general relativity:} various papers (Lee 2013\nocite{lee13}, Chamberlin \& Siemens 2012\nocite{cs12}, Lee et al. 2010\nocite{ljp+10},  Lee, Jenet \& Price 2008\nocite{ljp08}) have described how the angular correlation curve will change for different theories of gravity. \cite{ljp08} showed that differentiating between the curves from different theories of gravity would require a timing array containing up to a few hundred pulsars and is therefore beyond the scope of existing PTAs.
\end{itemize}

\subsection{Summary}

\begin{figure}
\begin{center}
\includegraphics[angle=-90,width=10cm]{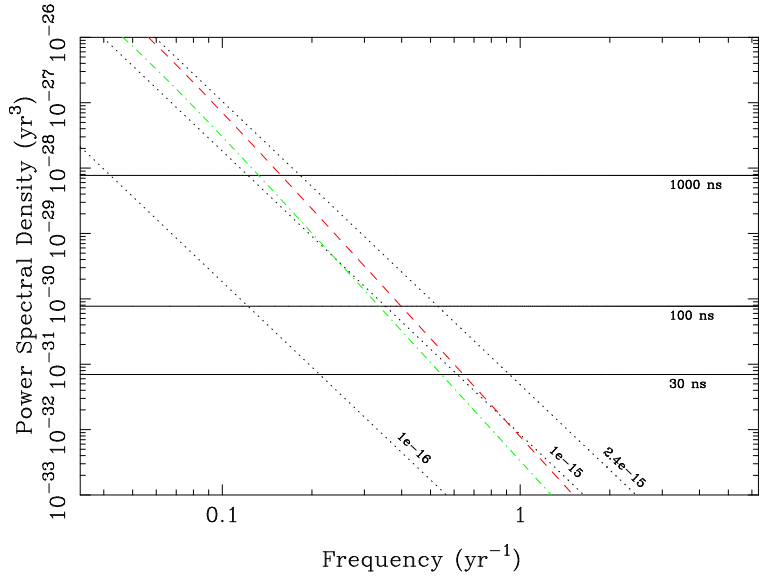}
\caption{Predicted power spectral density as a function of frequency for a background of GWs with $A = 2.4 \times 10^{-15}$, $A = 10^{-15}$ and $10^{-16}$ (dotted lines). Overlaid are the spectra of the expected errors in International Atomic Time, TT(TAI) (red, dashed line) and the JPL solar system ephemeris DE414 (green, dot-dashed line).  Power spectral density corresponding to white noise with rms values of 30, 100 and 1000\,ns with a 14 day cadence are shown as horizontal solid lines.} \label{fg:spectra1}
\end{center}
\end{figure}

In Figure~\ref{fg:spectra1} we show the power spectral density as a function of frequency for GWs, clock errors and planetary ephemeris errors.  The GW spectra density is calculated for a GW background of merging black hole binaries with $A = 10^{-16}$ and also for $A = 10^{-15}$. We also plot the current upper bound of $A = 2.4 \times 10^{-15}$ \citep{src+13}.  Irregularities in TT(TAI) with respect to TT(BIPM2013) are shown as the red, dashed line.  This line is representative, as 1) a given realisation of Terrestrial Time improves as new clocks are added (therefore showing the noise as a simple power law spectrum is slightly misleading), and 2) the stability of TT(BIPM) is expected to be significantly better than TT(TAI).  Predicting a similar curve for the expected instabilities in TT(BIPM) is currently ongoing and will be presented elsewhere.  The green dot-dashed line indicates the difference between the JPL planetary ephemerides DE421 and DE414.  Again, this line should be considered an upper bound on the residuals induced by the planetary ephemeris. 

Even though the exact level of the signals of interest are still uncertain, it is clear that all three effects (GWs, clocks and ephemeris errors) can contribute to the timing residuals, thereby increasing the complexity of extracting any one signal.  The indication from the figure that even a pulsar timed at the 1$\mu$s level would be significantly affected by these signals after $\sim 8$\,yr suggests that FAST should easily be able to make a detection.  However, as shown in the next section, other noise processes will make extracting these signals challenging.

\section{Prospects for high precision timing with FAST}\label{sec:limitNoise}

With the 64-m diameter Parkes telescope, a 256\,MHz bandwidth and observation times of $\sim$1\,hr  it is possible to achieve data sets with rms timing residuals $\sim$100\,ns  on only a few pulsars (see, e.g., Manchester et al. 2013\nocite{mhb+13}). Simply scaling this precision according to the radiometer equation suggests that FAST should be able to achieve $\sim 1-10$\,ns timing precision on at least some pulsars.  Unfortunately, as shown in this section, it is unlikely that this will be achieved.

Cordes \& Shannon (2010) and \cite{jat11} list various noise sources that will affect PTA data sets.  These include effects such as tropospheric fluctuations and mechanical noise.  Most of these effects are either correctable or at a small enough level to ignore over a reasonable data span.  In this section we consider the three major noise sources: jitter noise, dispersion measure variations and intrinsic timing noise.  We conclude the section by considering the effects of the polarisation calibration, radio frequency interference and the time distribution system.

\subsection{Jitter noise}

As part of the PPTA project 24 pulsars are currently observed.  The majority of observations for most of these pulsars are sensitivity limited and, hence, observing with a larger telescope would improve the precision and accuracy with which pulse arrival times could be determined.  However, as shown in \cite{ovh+11} observations of PSR~J0437$-$4715 do not improve with increased sensitivity.  This is due to ``jitter noise" which results from intrinsic variability in the shape of individual pulses from the pulsar\footnote{\cite{ovh+11} suggest that this phenomenon should be termed Stochastic Wideband Impulse Modulated Self-noise (SWIMS).}.  They showed that, even with a large telescope, the timing precision for this pulsar could never be better than $\sim 40$\,ns with a 1\,hour observation (and worse with shorter observations).  PSR~J0437$-$4715 is too far south for FAST to observe, but \cite{sc12} have found similar results using Arecibo observations of PSR~J1713+0747.  More recently \cite{sod+14} showed that, during bright scintillation states, seven of 22 pulsars in the PPTA exhibit jitter.   It is therefore likely that FAST observations of most of the currently known millisecond pulsars will be in the jitter dominated regime.  

For the following we emphasise that these are order-of-magnitude calculations.  Any individual pulsar will not exactly follow these predictions. In order to provide an initial estimate of how many pulsars will be jitter dominated with FAST we first estimate the jitter noise level for each pulsar (Shannon \& Cordes 2012):
\begin{equation}\label{eqn:jitter}
\sigma_{\rm J} \approx 0.2 W \sqrt{\frac{P}{t}}
\end{equation}
where all parameters are measured in seconds, $W$ is the pulse width, $P$ the pulse period and $t$ is the integration time.  In order to determine whether a specific pulsar will be, on average, jitter dominated or radiometer-noise dominated we calculate the expected noise level for the radiometer noise using:

\begin{equation}
\sigma_{\rm rad.} \approx \frac{W}{\rm S/N} \approx \frac{W T_{\rm sys}}{GS\sqrt{2\Delta ft}}\sqrt{\frac{W}{P-W}}
\end{equation}
where S/N is the profile signal-to-noise ratio, $T_{\rm sys}$ is the system temperature, $G$, the telescope gain, $S$ the pulsar's flux density and $\Delta f$ the usable bandwidth.  In Table~\ref{tb:fastPsrs} we tabulate (for all pulsars with a known pulse width and flux density) whether each pulsar would be jitter dominated with FAST (assuming nominal parameters of $G=16.5$\,K\,Jy$^{-1}$, $T_{\rm sys} = 20$K, $\Delta f = 800$\,MHz), a telescope similar to the Parkes telescope with existing receivers ($G=0.8$\,K\,Jy$^{-1}$, $T_{\rm sys} = 28$K, $\Delta f = 256$\,MHz) and for a telescope similar to the proposed Qitai design ($G=2.4$\,K\,Jy$^{-1}$, $T_{\rm sys} = 20$K , $\Delta f = 2000$\,MHz)\footnote{The Xingjiang Qitai 110m Radio Telescope (QTT) is a planned, fully-steerable single-dish telescope that will operate over a large frequency range.  It is not clear which telescopes in the Northern Hemisphere will still be operating in the FAST-era, but it is likely that a large number of 100-m class telescopes  (in Europe, China and North America) will continue to observe pulsars.}. The available bandwidths on the specified telescopes are likely to be larger than those assumed, however, here we are only carrying out order-of-magnitude estimates and we have chosen a smaller bandwidth to account for effects caused by changes in the sky temperature as a function of observing frequency, radio-frequency interference, scattering effects and the pulsar spectral indices.  In the final three columns of the table we list the time predicted for FAST to reach a timing precision of 100\,ns and 30\,ns along with the timing precision achievable after 15 minutes. 

For the Parkes-style telescope only one pulsar in the sample (PSR J2145$-$0750) is expected to be jitter dominated for the majority of observations. However, as pulsars scintillate they can become significantly brighter.  Therefore bright observations of some of these pulsars are jitter dominated, whereas typical observations may not be.  For the Qitai-style telescope it is likely that six of the 18 pulsars will be jitter dominated.   For FAST all except two will be in the jitter regime.  Therefore no advantage is obtained (with respect to timing precision) for observing PSRs J1022$+$1001,  J1713$+$0747, J1939$+$2134 or J2145$-$0750 with FAST instead of with a smaller telescope such as Qitai or other similar Northern hemisphere telescopes.

  This leads to:
\begin{itemize}
\item Even with a very large telescope, long observation times $\sim 1$\,hour per pulsar will be required to obtain high time-precision observations for most of the currently known pulsars.  
\item For pulsars that are jitter dominated there is no advantage (for timing purposes) in using a very large telescope.
\item Predictions for the expected GW detection significances that assume that FAST will be able to obtain very high timing precision ($\sim 10$\,ns) are likely to be erroneous.
\end{itemize}  

Jitter noise is currently thought to be a limiting noise process. However, \cite{ovdb13} demonstrated an improvement of nearly 40\% in the rms timing residual for PSR J0437$-$4715 using information about the polarised pulse profile.  They argued that result was currently limited by variable Faraday rotation in the Earth's ionosphere.  Current studies of jitter noise are limited by the sensitivity of available telescopes.  Observations of pulsars with FAST will allow us to significantly improve our understanding of the jitter phenomenon. This improved understanding may allow the development of strategies to remove the effect of jitter both for FAST, the SKA and for smaller telescopes.

In Section~\ref{sec:realistic} we provide a description of the properties of the pulsars that could be observed by FAST and would optimise the role of FAST as part of the international PTA projects.

\subsection{Dispersion measure variations}

\begin{figure}
\begin{center}
\includegraphics[angle=-90,width=10cm]{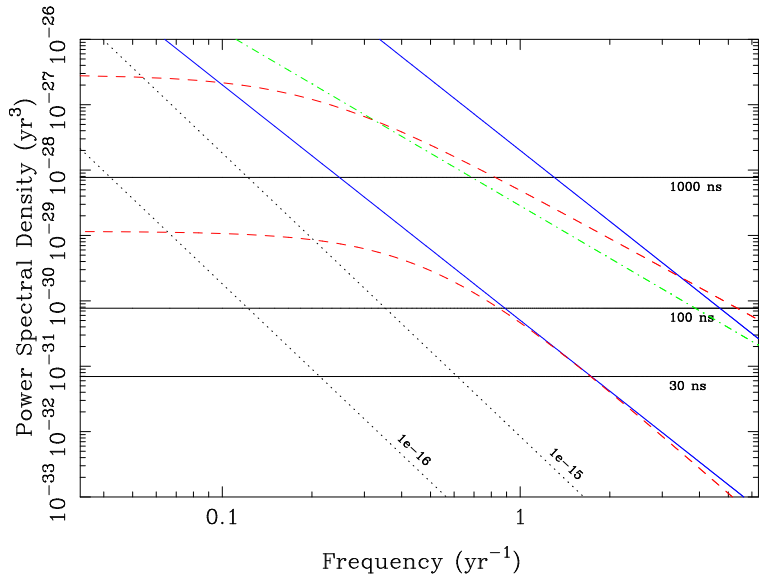}
\caption{Predicted power spectral density as a function of frequency for a background of GWs with $A = 10^{-16}$ and $10^{-15}$ (dotted lines). Overlaid are the spectra of the noise induced in the 20\,cm observing band by typical dispersion measure variations (green dot-dashed line). We overlay two representative timing noise models for PSRs J1939+2134 and J1909$-$3744.  The blue lines are the predictions for the red noise in these pulsars from Shannon \& Cordes (2010); the top line is for J1939+2134 and the lower line for J1909$-$3744.  Actual estimates of the noise for these pulsars from the Parkes pulsar data sets is given as red dashed lines. Power-spectral-density corresponding to white noise at 30, 100 and 1000\,ns with a 14 day cadence are shown as horizontal solid lines.} \label{fg:noise}
\end{center}
\end{figure}

In Figure~\ref{fg:noise} we show the power spectral density expected for a GW background at $A = 10^{-16}$ and $A = 10^{-15}$.  We overlay the expected spectrum for observations at 20\,cm affected by DM variations caused by Kolmogorov turbulence in the interstellar medium (green dot-dashed line).  The noise level is significantly higher than the GW background signal and therefore the effect of DM variations must be removed from the data sets.  \cite{kcs+13} provide a method to remove the DM variations which requires observations that are close in time at widely separated observing frequencies. As shown in the previous section, FAST will be affected by jitter noise for many pulsars.  If the observations in the low and high frequency bands are separated in time (i.e., if FAST initially observes with the 300\,MHz to 1.7\,GHz receiver and then subsequently observes with the 2 to 3\,GHz receiver) then both observations are likely to be jitter dominated thereby restricting the accuracy that the DM variations can be determined.  

In contrast, it is likely that FAST will obtain profiles with large S/N across the 300\,MHz to 1.7GHz band.  If jitter is a broad-band phenomena, such an observation should provide a DM determination with much better precision than is required to correct the resulting (jitter-dominated) arrival times\footnote{We note that multi-path scattering effects may limit the availability of the lower part of the band for determining DM variations that can be applied to arrival times determined from the high-frequency end of the band.}.

\subsection{Intrinsic timing noise}

\cite{hlk10} analysed timing irregularities for 366 pulsars allowing the first large-scale analysis of timing noise over time-scales of $>$10\,yr. The youngest pulsars were shown to be dominated by the recovery from glitch events. The timing irregularities for older pulsars seemed to exhibit quasi-periodic structure.  Millisecond pulsars were included in the sample, but the data set used (from the Lovell telescope at Jodrell Bank) did not allow an analysis of timing noise at the level necessary for PTA-research.  The first major assessment of timing noise on the precision timing of millisecond pulsars was given by \cite{sc10}.  They concluded that timing noise is present in most millisecond pulsars and will be measurable in many objects when observed over long data spans. They present simple models for predicting the amount of noise in a given pulsar.  In Figure~\ref{fg:noise} we overlay the \cite{sc10} predictions for PSRs J1939+2134 and J1909$-$3744 (blue solid lines - the higher line corresponds to PSR J1939+2134).  These models assume that the timing noise follows a simple power law with exponent $-3.6$.  Red noise models have been made for the actual PPTA data for these pulsars to whiten the residuals (these are overplotted as red dashed lines).  We note that these models were never made to be predictive.  However, the modelling suggests the possibility of a low-frequency turnover in the power spectra that implies that the power in the red noise plateaus at some level.  If both of these are true, this could make detecting GWs significantly easier.   Clearly knowing the frequency of turnover in the spectrum at which the noise level plateaus is essential in determining if the GW signal could be detected.  This plateau has physical meaning for models of timing noise based on superfluid turbulence \citep{ml14} and for those based on reflex motion from an asteroid belt \citep{scm+13}. 

There is currently no proven method for removing timing noise.  Timing noise will therefore provide a stringent limit for long-term timing projects.  FAST will need to either observe pulsars that exhibit small amounts of timing noise, observe a sufficiently large number of pulsars to achieve the science goals before timing noise dominates the residuals or observe over long enough data spans for the timing noise level to plateau\footnote{In this latter case a large telescope such as FAST is not needed and identical results could be obtained from a smaller telescope.}.  \cite{lhk+10} showed that the timing noise seen in the residuals for young pulsars can be modelled as a process in which the spin-down rate of the pulsar flips between two stable states. The pulse profile is seen to be different in the two states.  This leads to the possibilities that 1) millisecond pulsar timing noise can also be modelled as a two-state process and that 2) the state at a given time can be determined from the pulse shape.  If true, it may be possible to completely remove the effect of timing noise.

One IPTA pulsar, PSR~J1824-2452A, has been observed to undergo a small glitch event \citep{cb04}.  However, such events are rare (no other glitch event has been reported in any millisecond pulsar) and can be modelled with a simple rotation frequency step. It is therefore unlikely that glitch events will significantly restrict a future PTA.

\subsection{Calibration}

Millisecond pulsar profiles have complex morphology comprising many features.  Some of these components can be highly polarised whereas others may be unpolarised.  Inaccurate polarisation calibration can therefore lead to significant changes in the apparent pulse shape, hence leading to biases in the measured ToAs.  It will be essential to enable FAST observations to be calibrated with sufficient accuracy so that the ToAs are not affected at more than the $\sim$10\,ns level. 

In the latest in a series of papers, \cite{van13} described how polarimetric calibration can be carried out for pulsar observations.  It is necessary first to characterise the receiver being used.  For instance, this includes measuring the amount of cross-coupling between the feeds. Generally such parameters do not significantly change over months or even years.  However, it is also necessary to correct for differential phase and gain variations.  These can vary on timescales comparable to a standard pulsar observation.   Using observations of a pulsar with very high S/N, \cite{van13} showed how full calibration can be achieved if the polarisation profile of the pulsar is well known and is stable over time.  If such a pulsar is not available then it is necessary to use a system such as a pulsed calibration signal injected into the feed.  For the most precisely timed pulsars such calibrations are carried out at the Parkes observatory before and after each observation.  Absolute flux density calibration is not necessary for high precision pulsar timing, but does allow a large number of secondary science goals to be undertaken with the observations.  Flux density calibration is generally carried out by observing a standard source of known flux density. 

Over time, new backend instruments will be commissioned for FAST along with changes to the infrastructure such as cabling and the time distribution system.  Any such changes will lead to step-changes in the pulse arrival times which are usually dealt with  by including arbitrary phase jumps at the known times of the changes in the pulsar timing model.  It will therefore be necessary that any changes in the system are logged and advertised to the pulsar timing community. One possibility, currently being tested at Parkes (see Manchester et al. 2013\nocite{mhb+13}), is to measure precisely the time delay through the system after every change.  If known, this time delay can be used to remove the effect of any changes in the system and allow the pulse arrival times to be referred to a fixed point on the telescope.

\subsection{Effect of radio interference}

Even though FAST is being built within a karst depression in a radio quiet zone \citep{nlj+11} the observations will be affected by terrestrial and space-borne radio frequency interference (RFI).   The timing precision with which FAST will measure pulse arrival times implies that even small amounts of RFI could significantly affect those arrival times.  RFI causes arrival-time fluctuations by affecting 1) the ability to calibrate the data and 2) modifying the shape of the resulting pulse profile.  Over the wide observing bands used there will be parts of the band that are continuously affected by strong RFI, other parts that will be intermittently affected (e.g., by aircraft or satellites) and other regions that will be relatively clean.  It will be necessary to ensure, when designing the receiver system, that RFI in one part of the band cannot affect any other part.  

Various methods have been proposed to remove the effects of RFI.  These include the use of an automatic removal procedure based on spectral kurtosis (see Nita \& Gary 2010\nocite{ng10}).  However, such techniques will have limited use on FAST as the signal to noise of individual pulses may be larger than one.  Such methods would therefore clip the pulses leading to distorted pulse shapes.  The low frequency observing band at the Parkes observatory is affected by digital television channels.  \cite{khc+05} showed how adaptive filters can be used to remove the RFI without affecting the underlying pulsar signal.  This method relies on the RFI having a strong signal in a reference antenna and requires that the FAST backend instrumentation simultaneously process signals both from FAST and from the reference antenna.

RFI clearly affects the determination of ToAs. However, an in-depth study of RFI on the timing of millisecond pulsars has not yet been carried out and a large number of mitigation strategies exist.  It is therefore currently hard to identify exactly how RFI will affect FAST data sets and whether it will become a limiting noise process.

\subsection{The time distribution system}

In order to carry out high precision pulsar timing, it will be necessary for pulse ToAs to be referred to a realisation of TT.  This is not trivial and will need to be carried out with care in order to ensure that any variations in the time signal used for the determination of ToAs (at the nanosecond level) are recorded.  Our current software accounts for clock variations and the time transfer using a set of clock correction files.  As an example, the time transfer at Parkes can be obtained in two different ways (see Manchester et al. 2013 for details).  One system uses a global positioning system (GPS) clock which directly gives UTC(GPS)-UTC(PKS) at 5 minute intervals.  The BIPM publishes TT(TAI)-UTC(GPS) and so the time transfer from Parkes to TT(TAI) is known.  The second system uses a GPS common-view link to UTC(AUS).  From there it is possible to convert directly to TT(TAI).  Whatever system is used, it will be necessary for the instrumentation at FAST to tag pulse arrival times precisely and to have a way to convert the observatory clock to a realisation of Terrestrial Time.

\section{A realistic PTA}\label{sec:realistic}

\begin{figure}
\begin{center}
\includegraphics[angle=-90,width=12cm]{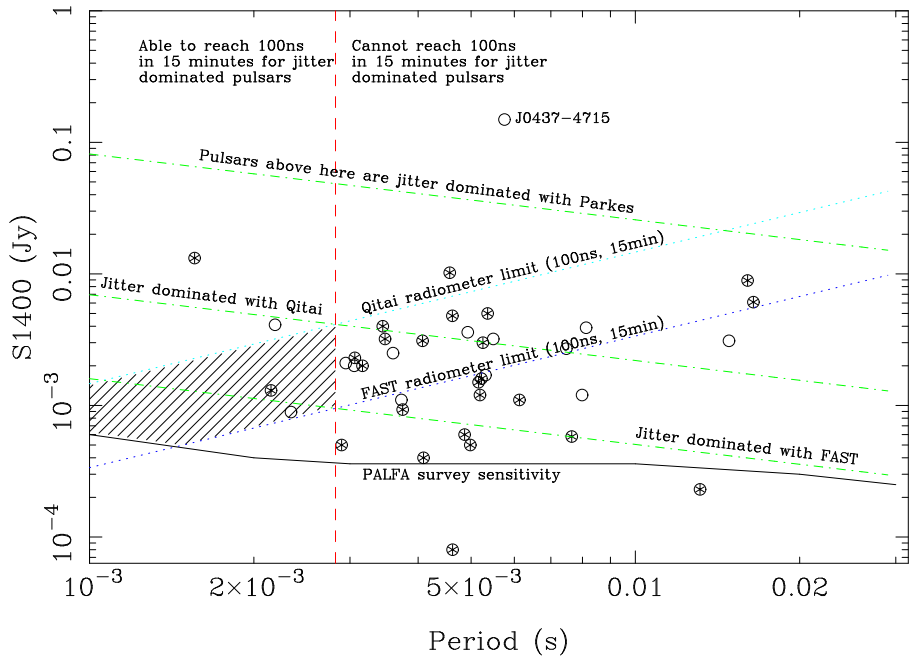}
\caption{Flux density versus pulse period parameter space. Regions for which pulsar observations will be dominated by jitter noise or radiometer noise are delimited. In all calculations a pulse duty cycle of 10\% is assumed.} \label{fg:fastPsrsSelect}
\end{center}
\end{figure}

It seems plausible that FAST will be able to reach the jitter noise level for most of the currently known pulsars. The final columns of Table~\ref{tb:fastPsrs}  imply that, for most of these pulsars, FAST will require significant observation durations in order to achieve timing precision of interest for PTA research.  An example of the type of pulsar that FAST will significantly improve over smaller telescopes is PSR J1741$+$1351.  This is a fast spinning pulsar with a narrow profile.  FAST will be jitter dominated, but other smaller telescopes will not be. A telescope such as Parkes would require over 180 hours to achieve a timing precision of 100\,ns.  Qitai would require 1.4 hours and FAST would reach the same timing precision in 10 minutes.

A reasonable, and realistic, timing array on FAST would include observations of around 50 pulsars that can achieve timing precisions of $\sim 100$\,ns.  For a realistic amount of observing time it seems likely that all these pulsars would need to be observed within $\sim$24\,hours. This gives observation times per pulsar of around 15 minutes (allowing for time to slew between sources and carry out calibration observations).    We can estimate the properties of such pulsars by assuming a duty cycle\footnote{the mean duty cycle for the pulsars currently observed for the IPTA is 0.09.  However the standard deviation is 0.07.} of 10\%\ and using the parameters for the gain, system temperature and bandwidth for the telescopes as considered earlier.  We again emphasise that these are simple estimates and any given pulsar will not necessarily follow the predictions of this calculation.  

In Figure~\ref{fg:fastPsrsSelect}, we plot the pulse period versus flux density in the 20\,cm band for all the IPTA pulsars for which such information exists (open circles).  Pulsars that lie within the FAST sky coverage are highlighted with star symbols.  Using the assumptions given above we can draw  lines on the Figure that, for a pulsar observation dominated by radiometer noise, delimit pulsars that could be timed with a ToA precision of 100\,ns in 15 minutes (those above the line).  Such lines are shown for Qitai and for FAST.  It will never be possible to obtain such timing precision for pulsars below these lines without longer integrations or by improving the observing system (such as increased bandwidth).  Therefore for around 15 pulsars in the current IPTA it would not be possible for FAST to obtain the precision that we require in the available observing time (and equivalently the Qitai telescope would not be able to achieve such precision for most of the current pulsars without longer integrations).

We can determine the regions on Figure~\ref{fg:fastPsrsSelect} in which pulsars would be jitter dominated for Parkes (top green, dot-dashed line), Qitai (central line) or FAST (bottom line).   The vertical red line in Figure~\ref{fg:fastPsrsSelect} delimits pulsars (on the left) which, if jitter dominated, would lead to a timing precision of 100\,ns or better within an observing time of 15 minutes. We therefore identify a region, the shaded region, in which pulsars observed by FAST could obtain the necessary timing precision in the available time. This shaded region is deliminated by the FAST radiometer limit, the jitter limit and a representative survey sensitivity (we use the sensitivity of the PALFA survey with Arecibo; Cordes et al. 2006\nocite{cfl+06})\footnote{ For completeness we note that the pulsar with the smallest flux density in the Figure is PSR J1911$+$1347.  This was discovered in the Parkes multibeam survey \citep{fsk+04} and has a flux density significantly lower than the nominal sensitivity of that survey.}.  The upper bound of the shaded region is deliminated by the Qitai radiometer limit. For pulsars above that line FAST could also obtain the required timing precision, but the Qitai telescope would also be able to provide identical results.
 
 Currently there is one pulsar in the shaded region that would be detectable using FAST (PSR J1903+0327) and one that is too far South (PSR J1017$-$7156). These pulsars were discovered in recent surveys using the Arecibo telescope (Champion et al. 2008\nocite{crl+08}) and Parkes \citep{kjb+12} respectively.  With improved calculations (based on, for instance, true pulse widths) it is likely that $\sim$10 known pulsars would lie within the shaded area.  It is likely that current Northern hemisphere pulsar surveys (for instance see Boyles et al. 2013\nocite{blr+13} and Ng 2013\nocite{ng2013}) will provide a few more pulsars  before FAST becomes operational.  However, to succeed in timing 50 pulsars at 100\,ns timing precision it will be necessary for FAST to carry out pulsar surveys that are sensitive to fast (\lapp3\,ms), narrow (with duty cycles \lapp10\%) pulsars with flux densities around $1$\,mJy.  Predicting whether this is possible is challenging and a description of pulsar surveys will be presented elsewhere in this series of articles. However, we note that PTA-style pulsars can be found in traditional large-scale pointed surveys, in drift scan surveys or even through observations of selected regions of sky (for instance, many of the Fermi gamma-ray sources have been found to be millisecond pulsars).

In order to produce the data sets described above, it will be necessary to discover a large number of pulsars using FAST and to carry out standard pulsar timing programs to identify which pulsars are jitter dominated and which provide PTA-quality arrival time measurements.  Carrying out the PTA observations, discovering new pulsars and carrying out follow-up observations of the newly discovered pulsars will require a significant fraction of the available observing time on FAST.

\subsection{Data archiving, storage and processing}\label{sec:archive}

Traditionally pulsar astronomers have not considered data archiving for PTAs a major problem.  However, as data rates increase, recording, archiving and processing are all becoming more challenging. Typical data file sizes for a single pulsar observation are currently around $\sim 1$GB (for a data file containing 64 subintegrations, 1024 frequency channels and 2048 phase bins).  With wider bands available it is plausible that FAST would produce data files of $\sim 10$\,GB for each standard observation. During a year of observing, 50 pulsars with weekly cadence will lead to $\sim 30$\,TB for each backend instrument\footnote{The PPTA team are finding it useful to have multiple backends recording the same data.  This enables determination of instrumental effects that are otherwise hard to identify.}.  This is not impractical, but it will be necessary to provide copies of these data at the data processing centres.  It is likely that new methods will soon be developed to improve pulsar timing.  These could include studies of small variations in pulse shape, timing using bright individual pulses, using higher-order moments of the electric field or making use of dynamic spectra. Such analysis methods would significantly increase the data volumes to process and store. The maximum data rate possible would arise from recording baseband-sampled data.  For the initial receiver design this would lead to (for 8-bit sampled data) $\sim 6$\,GB/s. For a typical 15 minutes observation this would lead to a data file of 5\,TB and a total volume per observing session for 50 pulsars of 245\,TB.  Making use of an RFI mitigation method may also significantly increase the data storage requirements.

The supporting infrastructure that enables multiple copies of the data to be replicated at different sites and enables the processing of that data will need to be planned in detail.  Keeping these systems operating and providing data and processing power to the community is non-trivial and will require significant management. Recent work with data sets from the Parkes observatory has shown the enormous value in keeping well maintained, long term observational data collections (see Hobbs et al. 2011\nocite{hmm+11}) in a standard format (such as PSRFITS; Hotan, van Straten \& Manchester 2004\nocite{hvm04}\footnote{\url{http://www.atnf.csiro.au/research/pulsar/index.html?n=Main.Psrfits}}).  Making these data available to the international community (after an embargo period) will ensure maximum science return from the observations. For the PTA teams to make full use of the data sets it will be necessary for the data sets stored in the repository to contain not only the original observation files, but also information describing the experimental configuration and details of the algorithms performed on the files.  Storing sufficient information about the data set allows the results to be reproducible, allowing science results to be confirmed, and enables other, non-PTA, science projects to be carried out with the data.

Standard processing of the pulsar arrival times requires minimal compute resources.  A typical set of 260 observations for a given pulsar spread over five years can easily be processed with \textsc{tempo2} on a laptop computer. However, with the wide bandwidths available it will be likely that a large number of ToAs at different frequencies will be determined for a single observation. Many of the gravitational wave detection codes (both frequentist and Bayesian based methods) require the analysis of $N_{\rm toa} \times N_{\rm toa}$ matrices and many of the algorithms require $N_{\rm toa}^3$ operations.  This can quickly become prohibitive for a large number of ToAs, particularly if it is necessary to run the detection code on a large number of simulated data sets.  It is therefore essential that the FAST data sets can be processed on high performance computers.  Much of the pulsar processing software is now being modified to allow the use of Graphics Processing Units (GPUs) where available.  GPUs are well suited for dealing with processing large matrices and we therefore recommend that the infrastructure created for FAST allows the data to be processed both on Central Processing Units (CPUs) and also on GPUs.

\subsection{Possibilities for public outreach}\label{sec:outreach}

Any project that makes use of the world's largest telescope to study pulsars, black holes and gravitational waves will have huge, international public interest.  Pulsar timing experiments do not fit well into citizen science projects such as \emph{Galaxy Zoo} because they require experienced astronomers to process the data\footnote{In contrast pulsar surveys are suited to citizen science projects. See, for example, the Einstein@Home project; Knispel et al. (2010)\nocite{kac+10}.}.  However, the pulsars observed are bright and (particularly with a large telescope such as FAST) can easily be detected within a few minutes.  This leads to the possibility of a PULSE@Parkes-style  outreach project (see Hobbs et al. 2009\nocite{hhc+09}) in which high school students around the world carry out observations of pulsars using the Parkes radio telescope.  Currently over 1000 students (from 110 schools) have taken part in this project from Australia, the Netherlands, USA, Japan, England and Wales.  Professional astronomers gain extra observations of their pulsars.  The students gain experience in observing with a large radio telescope and carry out small experiments based on their observations (such as measuring a pulsar's dispersion measure). The inclusion of FAST and the involvement of researchers in China in such projects would significantly increase the outreach potential of pulsar astronomy.

\section{Research questions relating to FAST and timing array research}\label{sec:unanswered}

There are a large number of open research questions relating to pulsars and timing array projects.  Here we list some of the most important whose solutions could affect the design of a PTA for FAST.

\begin{itemize}
\item \emph{Time standards:} we can easily study the difference between TT(TAI) and TT(BIPM), but we would really like to know the expected errors in TT(BIPM).  The expected stability of the various clocks used in forming TT(BIPM) is known, but has not yet been presented in a way to predict the timing precision needed for a PTA to detect such variations.

\item \emph{Solar system ephemerides:} out of the three solar system planetary ephemerides, the EPM ephemeris is the most divergent. Unfortunately this ephemeris currently cannot be used within the \textsc{tempo2} software package. Updating the software to enable its use would allow researchers to attempt to distinguish between the models using PTA data sets.

\item \emph{Timing noise:} intrinsic pulsar timing noise could significantly limit the possibility for GW detection using pulsar timing arrays. It is therefore essential that an improved understanding of timing noise is obtained as soon as possible.  Questions of interest include 
 \begin{enumerate}
  \item can timing noise be parameterised by a simple red noise model?
  \item does the power in the timing noise continue to increase with longer data spans or is there a cutoff frequency at which the power in the noise plateaus?
   \item can the amount of noise be predicted from e.g., the pulsar spin parameters?
   \item is there a class of pulsar that exhibits significantly less timing noise than another class?  
   \item can timing noise be removed by, for instance, studying pulse shape state changes? 
  \end{enumerate}
  
\item \emph{Jitter:} it is expected that jitter will be a dominant noise process.  Questions include whether jitter is broadband across the FAST observing band from $\sim$700MHz  to $\sim$3GHz, whether the improvement that \cite{ovdb13} identified is applicable to other pulsars and how jitter will affect correction for DM variations.

\item \emph{Optimal observing strategy accounting for other telescopes:} FAST will be one of many telescopes worldwide undertaking observations of millisecond pulsars.  These include the existing telescopes that currently make up the IPTA, along with low frequency telescopes such as LOFAR and MWA, the SKA precursors and the SKA itself.  For the EPTA, \cite{lbj+12} have developed a method to optimise multi-telescope observing schedules in order to maximise the sensitivity to a GW background. For FAST the optimal observing schedule would depend on the ability to share data with other groups and a choice of the major scientific goals for the PTA (e.g., GW detection, or studying an already detected signal).

\item \emph{How long will it take to detect GWs?} This key question is still unanswered for data sets with realistic sampling and noise processes. It also depends upon the expected signal which itself is poorly known. An answer would provide an indication as to whether FAST will be attempting to make a first detection, or will be studying previously discovered GWs.
\end{itemize}

\section{Conclusion}

FAST will be an ideal telescope for participating in PTA projects.  It is possible that observations using FAST alone would lead to very exciting results such as the first direct detection of ultra-low frequency GWs, the identification of irregularities in terrestrial time scales or the discovery of a new object in the solar system.  However, the goals described here are the primary aims of the existing International Pulsar Timing Array project.  Combining the initial FAST data sets with the much longer-baseline IPTA data sets would be of huge benefit to both projects.  On a longer time scale combining the FAST data with observations from the SKA will produce ideal data sets for PTA research.  Even if it takes longer than expected for the main PTA goals to be reached, the data sets obtained by FAST will be used for numerous science projects including studies of the interstellar medium, planetary navigation and tests of theories of gravity. PTA science on FAST will enable Chinese scientists to become involved in cutting edge research that links neutron stars, black holes, galaxies, planets, gravitational waves and clocks whilst providing huge technical, statistical and computational challenges.

\section{Acknowledgements}

GH acknowledges support from the Australian Research Council (ARC) Future Fellowship programme.  SD acknowledges support from the National Natural Science Foundation of China (11225314).  We thank J. Zic and S. Os{\l}owski for comments on the manuscript. 
\bibliography{refs.bib}{}
\bibliographystyle{raa}

\label{lastpage}

\end{document}